\documentclass[12pt,showpacs,showkeys]{revtex4}
\usepackage{amssymb,amsmath}
\usepackage{epsf,epsfig}

\def\a{\alpha}
\def\b{\beta}
\def\g{\gamma}
\def\m{\mu}
\def\n{\nu}
\def\vt{\vartheta}
\def\stareq{\stackrel{*}{=}}

\begin{document}

\title{To consider the electromagnetic field as fundamental, and the metric
only as a subsidiary field}

\author{Friedrich W. Hehl}
\email{hehl@thp.uni-koeln.de}

\affiliation{Institute for Theoretical Physics, University of Cologne, 50923
  K\"oln, Germany \\ and\\Department of Physics and Astronomy, University of
  Missouri-Columbia, Columbia, MO 65211, USA}

\author{Yuri N. Obukhov}
\email{yo@thp.uni-koeln.de}

\affiliation{Institute for Theoretical Physics, University of Cologne, 50923
  K\"oln, Germany \\ and\\ Dept.\ Theor.\ Physics, Moscow State University,
  117234 Moscow, Russia}

\date{21 April 2004, {\it file forerunner5.tex}}
\bigskip

\begin{abstract}
  In accordance with an old suggestion of Asher Peres (1962), we
  consider the electromagnetic field as fundamental and the metric as
  a subsidiary field. In following up this thought, we formulate
  Maxwell's theory in a diffeomorphism invariant and
  metric-independent way. The electromagnetic field is then given in
  terms of the excitation $H=({\cal H},{\cal D})$ and the field
  strength $F=(E,B)$. Additionally, a local and linear ``spacetime
  relation'' is assumed between $H$ and $F$, namely $H \sim \kappa F$,
  with the constitutive tensor $\kappa$. The propagation is studied of
  electromagnetic wave fronts (surfaces of discontinuity) with a
  method of Hadamard. We find a generalized Fresnel equation that is
  quartic in the wave covector of the wave front. We discuss under
  which conditions the waves propagate along the light cone.  Thereby
  we derive the metric of spacetime, up to a conformal factor, by
  purely electromagnetic methods.
\end{abstract}



\maketitle

\section{Introduction}\label{intro}

``It is therefore suggested to consider the electromagnetic field as
fundamental, and the metric field only as a subsidiary quantity.''
This is a quotation from an article that Asher Peres \cite{Peres62}
wrote in 1962. Asher's idea was to start with the source-free Maxwell
equations
\begin{equation}\label{maxwell}
dH=0\,,\qquad dF=0\,;
\end{equation}
here we have the excitation $H=({\cal H},{\cal D})$ and the field
strength $F=(E,B)$. The equations (\ref{maxwell}) are diffeomorphism
invariant and independent of the metric $g_{ij}(x)$ of spacetime. In
other words, this is the {\it premetric\/} form
\cite{Einstein16,Kottler22,TT60,Post62,Richard65,Attay} of Maxwell's
equations (without source). The excitation $H$ and the field strength
$F$ are considered to be directly measurable quantities, via
Maxwellian double place in the case of $\cal D$ and via the Coulomb
force in the case of $E$; analogous procedures exist for $\cal H$ and
$B$, respectively, see \cite{birkbook}.

Peres then considered the ``constitutive'' law relating $H$ and $F$ in
vacuum, namely
\begin{equation}\label{constit1}
H=\lambda\,^\star F\,,
\end{equation}
as a {\it definition} of the metric; $\lambda$ is a universal constant
and $\star$ the (metric dependent) Hodge star operator. In components
the metric is displayed more explicitly,
\begin{equation}\label{constit2}
  \check{H}^{ij}=\frac{\lambda}{2}\,\sqrt{-g} \,
  g^{ik}g^{j\ell}\,F_{k\ell}\,,
\end{equation}
with $\check{H}^{ij}=\frac 12\,\epsilon^{ijk\ell}\, H_{k\ell} $, where
we used the totally antisymmetric Levi-Civita symbol
$\epsilon^{ijk\ell}=\pm 1,0$. The components of the excitation and
field strength 2-forms are given by $H=H_{ij}\,dx^i\wedge dx^j/2$ and
$F=F_{ij}\,dx^i\wedge dx^j/2$, respectively. The components
$\check{H}^{ij}$ and $F_{k\ell}$ are assumed to be known and
(\ref{constit2}) has to be resolved with respect to the metric
$g_{ij}$. This was the program of Peres \cite{Peres62} for deriving
the metric from the electromagnetic field.

Without doubt, Peres is the forerunner of the movement to construct
the metric of spacetime out of purely electromagnetic data. The
algebraic method used by Peres in his attempt was analyzed in detail
by Rubilar \cite{Guillermo02}. Subsequent to Peres \cite{Peres62},
Toupin \cite{Richard65} and Sch\"onberg \cite{Schonberg71} proved the
existence of a metric in this context.  Nowadays we know
\cite{birkbook} that the program, as layed out by Peres, can be
explicitly implemented. In this paper, which we would like to
dedicate to Asher Peres on the occasion 70th birthday, we are going to
sketch this procedure.

{\it Notation (see \cite{birkbook}):} We use the formalism of exterior
differential forms. We denote the frame by $e_\a$, with the
anholonomic or frame indices $\a,\b,\dots=0,1,2,3$.  Decomposed with
respect to a natural frame $\partial_i$, we have
$e_\a=e^i{}_\a\,\partial_i$, where $i,j,\dots=0,1,2,3$ are holonomic
or coordinate indices. The frame $e_\a$ is the vector basis of the
tangent space at each point of the 4D spacetime manifold. The symbol
$\rfloor$ denotes the interior and $\wedge$ the exterior product. Both
products are metric independent.  The coframe $\vt^\b=e_j{}^\b dx^j$ is
dual to the frame, i.e., $e_\a\rfloor \vt^\b=\delta^\b_\a$.

\section{Premetric electrodynamics}\label{premetric}

Since electric charges are occurring in nature in integer multiples of
$e/3$, here $e$ is the elementary charge, they can be counted.
Accordingly, if we consider a 3-dimensional (3D) volume, we can count
how many elementary charges are contained in it. Macroscopically in 4
dimensions (4D), we can describe the charge density and its flux by
the 3-form $J$ that is conserved:
\begin{equation}\label{charge}
dJ=0\,.
\end{equation}
If the global version of charge conservation (\ref{charge}) is
suitably formulated, by de Rham's theorem the electric current $J$
turns out to be exact:
\begin{equation}\label{inhMax}
J=dH\,.
\end{equation}
This is the inhomogeneous Maxwell equation, with the excitation 2-form
$H$.

In deriving (\ref{charge}) and (\ref{inhMax}), only the ability to
define an arbitrary 4D volume is necessary together with the counting
of electric charges. No distance measurement nor any parallel transfer
is involved. Therefore it is evident that the inhomogeneous Maxwell
equation already exists on a 4D manifold without metric and without
linear connection.

With the help of the Lorentz force density
\begin{equation}\label{axiom2}
  f_\alpha=(e_\alpha\rfloor F)\wedge J\,,
\end{equation} 
we can define the electromagnetic field strength 2-form $F$. If $F$ is
integrated over a 2D area, it describes the magnetic flux lines
piercing through this area.

In certain situations, inside a superconducting region of a
superconductor of type II, magnetic flux lines are quantized and can
be counted. Here again, the counting of certain units, now the
magnetic flux quanta, together with the ability to circumscribe a 2D
area element, is all what is needed to formulate a conservation law.
This conservation law of magnetic flux we assume to be generally
valid:
\begin{equation}\label{axiom3}
dF=0\,.
\end{equation}
Again, no metric and no linear connection of spacetime is involved. As
we will discuss below, Faraday's induction law is a consequence of
(\ref{axiom3}) thereby giving (\ref{axiom3}) a firm experimental
basis.

Also without using a metric, we can decompose the fields entering the
Maxwell equations into 1 plus 3 dimensions \cite{birkbook},
\begin{equation}\label{dec1}
  H = -\,{\cal H}\wedge d\sigma + {\cal D}\,,\qquad J =\> -\,j\wedge
  d\sigma\, + \rho\,
\end{equation}
and
\begin{equation}\label{dec2}
  F = \quad E\wedge d\sigma +f_\lambda B\,,\qquad A = \quad -\varphi\,
  d\sigma\, + {\cal A}\,.
\end{equation}
Here $\sigma$ is the prototype of a time variable.  The Lenz factor we
put to one in accordance with the Lenz rule, $ f_\lambda=+1$, see
\cite{IH3} for details. We also decomposed the potential 1-form $A$
that is defined by $F=dA$.

Using these decompositions and substituting them into the Maxwell
equations (\ref{inhMax}) and (\ref{axiom3}), we find the conventional
form of the Maxwell equations,
\begin{equation}\label{evol1}
  dH=J\>\,\begin{cases}\>\, \underline{d}\,{\cal
      D}\>=\rho\,,
    \\ \quad \dot{{\cal D}}\,=\underline{d}\,{\cal
      H}-j\,,
\end{cases}\qquad 
 dF=0\>\,\begin{cases} \>\,\underline{d}\,B\>=0\,, \\ 
    \quad\dot{B}\,=-\underline{d}\,E\,,
\end{cases}
\end{equation}
cf.\ Sommerfeld \cite{Sommerfeld} and Scheck \cite{Scheck04}. The 3D
exterior derivative is denoted by $\underline{d}$, the time derivative
by a dot.

\begin{figure}
\epsfxsize=\hsize 
\epsfbox{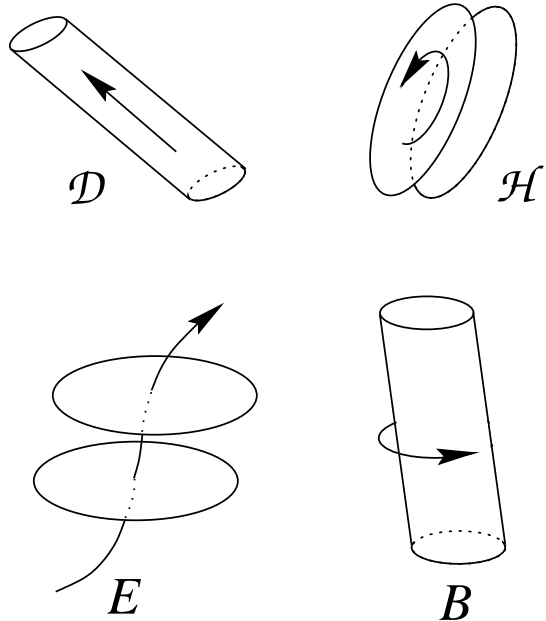}
\caption[{\it Faraday-Schouten pictograms of the electromagnetic field 
in 3-dimensional space. The images of 1-forms are represented by two
neighboring planes. The nearer the planes, the stronger the 1-form is.
The 2-forms are pictured as flux tubes. The thinner the tubes are, the
stronger the flow is. The difference between a twisted and an
untwisted form accounts for the two different types of 1- and 2-forms,
respectively.}]  {\label{schouten}{\it Faraday-Schouten pictograms of the
  electromagnetic field in 3-dimensional space. The images of 1-forms
  are represented by two neighboring planes. The nearer the planes,
  the stronger the 1-form is. The 2-forms are pictured as flux tubes.
  The thinner the tubes are, the stronger the flow is. The difference
  between a twisted and an untwisted form accounts for the two
  different types of 1- and 2-forms, respectively. Taken from Hehl and
  Obukhov \cite{birkbook}.}}
\end{figure}

We can complete this premetric electrodynamics by substituting
(\ref{inhMax}) into (\ref{axiom2}). Then, by some algebra and by some
physical arguments, one ends up with an axiom for the energy-momentum
current 3-form of the electromagnetic field:
\begin{equation}
   \Sigma_\alpha :={\frac 1 2}\left[F\wedge(e_\alpha\rfloor
    H) - H\wedge (e_\alpha\rfloor F)\right]\,.\label{simax}
\end{equation}

In equations (\ref{charge}) to (\ref{simax}), we collected all
fundamental relations of classical electrodynamics.  They are all
diffeomorphism invariant, that is, completely independent of the
coordinates used. With the exceptions of (\ref{axiom2}) and
(\ref{simax}), the equations are also invariant under arbitrary linear
frame transformations. In contrast, eqs.(\ref{axiom2}) and
(\ref{simax}), like the frame $e_\a$, transform linearly, that is,
they are {\it co\/}variant under linear frame transformations.

We started by counting electric charges and magnetic flux lines, used
no metric (i.e., no gravitational potential) and no linear connection
of spacetime, and found a generally covariant basis for
electrodynamics. In particular, there is no relation whatsoever to the
Poincar\'e group (also known as inhomogeneous Lorentz group). It
simply does not enter at that level of our setting up of classical
electrodynamics.  Classical electrodynamics is not so closely related
to special relativity as most representations in textbooks want to
make us believe.

The present framework can be suitably generalized in order to allow
for magnetic charges (tantamount of violating magnetic flux
conservation, i.e., $dF\ne 0$), see \cite{PhysLett04,K04}. However, since
magnetic charges have never been found experimentally, we will not
follow this train of thought any further.

In contrast, the violation of electric charge conservation, i.e.,
$dJ\ne 0$, would probably hit our axiomatic set-up fatally.  Firstly,
the inhomogeneous Maxwell equation (\ref{inhMax}) would be lost and
secondly, the Lorentz force density (\ref{axiom2}), with a
non-conserved charge, is no longer expected to qualify as a bona fide
defining equation for the electromagnetic field strength $F$.
L\"ammerzahl et al.\ \cite{Lam01,Lam04,Lam05} developed an
electrodynamic theory in the framework of which one can accommodate a
violation of electric charge conservation. They use it as a test
theory for interpreting corresponding experiments. They propose such
experiments in order to improve the experimental bounds for charge
conservation.

Since in elementary particle physics all evidence supports electric
charge conservation, we will stick to this principle and will continue
our considerations by assuming the validity of (\ref{charge}) and
(\ref{axiom3}), respectively.

One could ask why these two conservation laws are so ``stable''. The
absolute dimensions of current and field strength are $[J]=q\,$ and
$[F]=h/q$, respectively, with $q\sim$ dimension of charge and $h\sim$
dimension of action. Let us assume that the elementary charge $e$ and
Planck's constant $\hbar$ are really {\it constants\/} of nature, that
is, neither time nor space dependent.  Then the elementary quanta of charge
$e/3$ and flux $\pi\hbar/e$ are also constants, and this would assure
the permanence of charge and flux. The corresponding conservation
laws, including the premetric Maxwell equations (\ref{evol1}), are
thereby implied. The only problem with this argument is that flux
occurs in quantized form only under special circumstances. Closely
related ideas were put forward by Peres \cite{Peres02,Peres03}.

Many electromagnetic theories fulfill, indeed, electric charge and
magnetic flux conservation (\ref{charge}) and (\ref{axiom3}),
respectively. Let us quote as examples nonlinear Born-Infeld
electrodynamics \cite{BI34}, the quantum field theoretically motivated
pseudo-classical Heisenberg-Euler electrodynamics \cite{HE36}, and, as
a more recent case, the axion type electrodynamics of Carroll, Field,
and Jackiw \cite{Carroll90}. Itin \cite{Itin04} has shown that the
CFJ-electrodynamics can be put in premetric form. We just have to
assume the constitutive law $H=\,^\star F+a F$, with the
metric-dependent Hodge star $^\star$ and the scalar function
$a:=vt/2$, where $v$ is an absolute (i.e., prescribed) field and $t$
the time coordinate. A violation of the Lorentz symmetry is obtained
by postulating this explicit (and noncovariant) time dependence of the
constitutive tensor.

As we have seen, premetric electrodynamics turns out to be a useful
framework for classifying different models of electrodynamics.

\section{Local and linear spacetime relation}\label{strelation}

The premetric framework, which we discussed so far, is incomplete. As
we can read off from (\ref{evol1}), we have $3+3=6$ equations for
determining the time evolution of the electromagnetic field $(H;F)$.
However, the latter is described by the $6+6=12$ components $({\cal
  H}_a,\, {\cal D}_{ab}=- {\cal D}_{ba};\, E_a,\,B_{ab}=-B_{ba})$,
with $a,b,\dots=1,2,3$. Thus $6$ equations are missing.

Guided by what we know from ordinary Maxwell-Lorentz electrodynamics,
we will assume that the excitation $H$ is a functional of the field
strength $F$. For vacuum spacetime, such a relation is expected to be
{\it local}, i.e., no integrals are involved linking $H(x^i)$ to $F$
at another event (no heredity effects). Furthermore, linearity will be
assumed (linear response). Accordingly, a local and linear operator
$\kappa$ will be postulated to exist, that is, $H=\kappa(F)$. In
components $H=H_{\a\b}\,\vt^\a\wedge\vt^b/2$ etc., this {\it spacetime
  relation\/} reads,
\begin{equation}\label{loclin1}
  H_{\alpha\beta} = \frac{1}{2}\,\kappa_{\alpha\beta}{}
  ^{\gamma\delta}\, F_{\gamma\delta}\,.
\end{equation}
The constitutive tensor of spacetime $\kappa_{\alpha\beta}{}
^{\gamma\delta}$ can also be transformed into holonomic coordinates: $
\kappa_{\alpha\beta}{}^{\gamma\delta} = e^i{}_\alpha\, e^j{}_\beta\,
e_k{}^\gamma e_l{}^\delta \,\kappa_{ij}{}^{kl}$ (see
Schouten\cite{Schouten89}). Sometimes it is more convenient to apply a
constitutive tensor with all indices as superscripts. We raise two
indices with the help of the Levi-Civita symbol and find
\begin{equation}\label{chi}
\chi^{\a\b\g\delta}:=\frac 12\epsilon^{\a\b\m\n}
\kappa_{\m\n}{}^{\g\delta}\,.
\end{equation}

Alternatively, we can introduce a 6-dimensional vector space with the
collective indices $I,K,\dots=1, \dots,6 \equiv01,02,03,23,31,12$.
Then (\ref{loclin1}) can be rewritten as 
\begin{equation}\label{loclin2}
  H_I=\kappa_I{}^K\,F_K=\hat{\epsilon}_{IM}\chi^{MK}\,F_K\,.
\end{equation}
All information on the electromagnetic properties of spacetime is
encoded in the 36 components of $\kappa$ or $\chi$, respectively. It
is straightforward to decompose $\chi$ irreducibly under the linear
group $GL(4,R)$. The $6\times 6$ matrix $\chi^{IK}$ decomposes in its
symmetric tracefree, its antisymmetric, and its trace pieces:
\begin{eqnarray}\label{6ddecomp}
  \chi^{IK}&=&\left(\chi^{(IK)}-\frac
    16\,\epsilon^{IK}\chi_L{}^L\right)+\chi^{[IK]}+\frac 16\,
  \epsilon^{IK}\chi_L{}^L\,,\\ 36&=&\qquad\qquad
  20\qquad\qquad\oplus\>\;\, 15\>\;\, \oplus\qquad 1\,.\nonumber
\end{eqnarray}
Translated into the 4D formalism, we find
\begin{equation}\label{result1}
  \chi^{ijkl}={}
  \underbrace{^{(1)}\chi^{ijkl}}_{20,\,\text{principal}}
  +\underbrace{\epsilon^{ijm[k}\!\not\!S_m{}^{l]}
    -\epsilon^{klm[i}\!\not\!S_m{}^{j]}}_{15,\,
    \text{skewon}}+\underbrace{\epsilon^{ijkl}\,\alpha}_{1,\,
    \text{axion}}\,.
\end{equation}
We indicated the names and the number of independent components in the
last equation. In conventional Maxwell-Lorentz electrodynamics, only
the principal part $^{(1)}\chi^{ijkl}$ is assumed to contribute to the
spacetime relation. The tracefree $4\times 4$ matrix $\!\not\!S_i
{}^{j}$ describes the skewon part and the axial (or pseudo) scalar
$\a$ the axion part of $\chi$. Again, a transformation from
anholonomic coordinates can be achieved by the usual rule
$\chi^{ijkl}=e^i{}_\a\,e^j{}_\b\,e^k{}_\g
\,e^l{}_\delta\,\chi^{\a\b\g\delta}$.

We substitute (\ref{result1}) in the holonomic version of
(\ref{loclin1}). This yields the spacetime relation
\begin{equation}\label{crypto2a}  
  H_{ij}=\frac 12\,^{(1)}\kappa_{ij}{}^{kl}\,F_{kl}+2\, {\!\not
    \!S}_{[i}{}^kF_{j]k}+\a\,F_{ij}\,.
\end{equation} 
Still, the 6D version in $3+3$ form may be nearer to our intuition. It
reads, see \cite{birkbook} for details ($a,b=1,2,3$),
\begin{equation}\label{CR}
  \left(\begin{array}{c} {\cal H}_a \\ {\cal D}^a\end{array}\right) =
  \left(\begin{array}{cc} {{\cal C}}^{b}{}_a & {{\cal B}}_{ba} \\ 
      {{\cal A}}^{ba}& {{\cal D}}_{b}{}^a \end{array}\right)
  \left(\begin{array}{c} -E_b\\ {B}^b\end{array}\right)\,,
\end{equation}
with the constitutive tensors
\begin{equation}\label{kappachi}  
  \kappa_I{}^K=\left(\begin{array}{cc} {{\cal C}}^{b}{}_a & {{\cal
          B}}_{ba} \\ {{\cal A}}^{ba}& {{\cal D}}_{b}{}^a
    \end{array}\right)\,,\qquad \chi^{IK}= \left( \begin{array}{cc}
      {\cal B}_{ab}& {\cal D}_a{}^b \\ {\cal C}^a{}_b & {\cal A}^{ab}
    \end{array}\right)\,.
\end{equation}
For material media, such constitutive tensors are well-known, see
Lindell et al.\ \cite{Lindell} and Mackay and Lakhtakia
\cite{Mackay04}, for example.  Mackay and Lakhtakia consider a
material with electric/magnetic cross-terms, i.e., $ {\cal C}^a{}_b$
and $ {\cal D}_a{}^b $ are nonvanishing. (However, they require, see
also \cite{Lakhtakia04}, the axion part to vanish.) It is convenient
to write down explicitly the contributions of the principal, the
skewon, and the axion parts to the constitutive 3-matrices, cf.\ 
(\ref{result1}):
\begin{eqnarray}
  {\cal A}^{ab} &=& -\varepsilon^{ab} -
  \epsilon^{abc}\!\not\!S_c{}^0,\label{A}\\ {\cal B}_{ab} &=&\>\;
  \mu_{ab}^{-1} + \hat{\epsilon}_{abc}\!\not\!S_0{}^c,\label{B}\\ 
  {\cal C}^a{}_b &=&\>\; \gamma^a{}_b\, - (\!\not\!S_b{}^a - \delta_b^a
  \!\not\!S_c{}^c) + \alpha\,\delta_b^a,\label{C}\\ {\cal D}_a{}^b &=&\>\;
  \gamma^b{}_a\, + (\!\not\!S_a{}^b - \delta_a^b \!\not\!S_c{}^c) +
  \alpha\,\delta_a^b. \label{D}
\end{eqnarray}
The set of the two symmetric matrices
$\varepsilon^{ab}=\varepsilon^{ba}$, $\mu_{ab}^{-1} = \mu_{ba}^{-1}$
plus the traceless matrix $\gamma^a{}_b$ (with $\gamma^c{}_c =0$)
comprise the principal part of the constitutive tensor.  Usually,
$\varepsilon^{ab}$ is called {\it permittivity} tensor and
$\mu^{-1}_{ab}$ {\it reciprocal permeability\/} tensor
(``impermeability'' tensor), since they describe the polarizability
and magnetizability properties of a medium, respectively. The
cross-term $\gamma^a{}_b$ is related to the Fresnel-Fizeau effects. The
skewon contributions in (\ref{A}) and (\ref{B}) are responsible for
the electric and magnetic Faraday effects, respectively, whereas
skewon terms in (\ref{C}) and (\ref{D}) describe optical activity.

\section{Propagation of waves: Fresnel-Hadamard approach}\label{hadamard}

Wave propagation is a very important physical phenomenon in classical
field theory. In electrodynamics, one usually distinguishes between
physical optics and geometric optics. We will confine our attention to
the latter case. The appropriate formalism is then provided by the
Fresnel-Hadamard approach in which a wave is described in terms of the
propagation of a discontinuity of the field. Let us define the surface
of discontinuity $S$ locally by a function $\Phi$ such that $\Phi=
const$ on $S$. As usual, we denote by $\left[{\cal F}\right](x)$ the
discontinuity of a function ${\cal F}$ across $S$, and $q:=d\Phi$ is
the wave covector.  Then for an ordinary wave, the geometric Hadamard
conditions are satisfied across $S$:
\begin{eqnarray}\label{jump}
[H] &=& 0\,,\qquad [dH]=q\wedge h, \\
{[}F{]} &=& 0\,,\qquad [dF]=q\wedge f.
\end{eqnarray}
The 2-forms $h$ and $f$ describe the jumps of the derivatives of the
electromagnetic fields. Using Maxwell's equations (\ref{inhMax}) and 
(\ref{axiom3}), we find 
\begin{equation}
q\wedge h=0,\qquad q\wedge f = 0.\label{qwedgeh}
\end{equation}
The latter equation can evidently be solved by $f=q\wedge a$, with an
arbitrary covector $a$. Now we use the spacetime relation
$H=\kappa(F)$. The corresponding relation for the jump 2-forms reads
$h = \kappa(f)=\widetilde{\kappa} (f)+\alpha f$, where we separated
the axion piece and denoted the rest of the constitutive relation
(\ref{crypto2a}) by the $\widetilde{\kappa}$.  Substituting this into
(\ref{qwedgeh}), we finally obtain
\begin{equation}
q\wedge  h= q\wedge\widetilde{\kappa}(q\wedge a)=0\,.\label{qwedgeh1}
\end{equation}
The last equation can be considered as a kind of Fourier
 transform of the tensor analytical formula
 $\partial_\b(\widetilde{\chi}^{\a\b\g\delta}\partial_\g A_\delta)=0\,,$
 see Post \cite{Post62}, Eq.(9.40).

\subsection{Generalized Fresnel equation and Tamm-Rubilar tensor}\label{trt}

Now we are in a position to derive the Fresnel equation for the wave 
covector $q$. As a preliminary step we recall that a 3-form in 4D 
has four components. Consequently, we can conveniently recast the 
3-form equation (\ref{qwedgeh1}) into an equivalent 4-form equation by 
multiplication with $\vt^\a=e_i{}^\a dx^i$. Recalling that $a=a_\b\vt^\b$,
we find
\begin{equation}
q\wedge\vt^\a\wedge\widetilde{\kappa}(q\wedge\vt^\b)\,a_\b = q\wedge\vt^\a
\wedge{\frac 12}\widetilde{\kappa}_{\mu\nu}{}^{\g\b}q_\g\, a_\b\,\vt^\mu
\wedge\vt^\nu=0\,.\label{qwedgeh2}
\end{equation}
This algebraic system for the covector $a$ obviously admits the gauge
freedom $a_\b\rightarrow a_\b+q_\b\varphi$ with an arbitrary function
$\varphi$. In order to deal with this problem, we choose the first leg
of the local anholonomic coframe as $\vt^{\hat{0}}\stackrel{*}{=}q$.
This can always be done without restricting the generality of our
consideration. Then (\ref{qwedgeh2}) reduces to
\begin{equation}
  {\frac 12}\widetilde{\kappa}_{\mu\nu} {}^{\hat{0}\b}a_\b\,
 \vt^{\hat{0}}\wedge\vt^\a\wedge\vt^\mu\wedge\vt^\nu \stareq {0}\quad 
 \Rightarrow\quad \frac 12\epsilon^{\hat{0}acd}\,
  \widetilde{\kappa}_{cd}{}^{\hat{0}b}\,a_b\stareq {0}\,
\end{equation}
or, equivalently, to the system of three algebraic equations for the
three unknown components $a_b$:
\begin{equation}
  W^{ab}a_b\stareq \hat{0}\,\quad{\rm with}\quad W^{ab}:=\widetilde{\chi}
  ^{\hat{0}a\hat{0}b}\,.
\end{equation}
Note that the gauge-dependent $a_{\hat{0}}$ eventually disappeared. A
nontrivial solution of the system obtained exists if and only if the
corresponding determinant vanishes,
\begin{equation}\label{fresnel1}
  {\cal W}:=\det W^{ab} \stareq
  \frac{1}{3!}\hat{\epsilon}_{abc}\hat{\epsilon}_{def}
  W^{ad}W^{be}W^{cf}\stareq 0  
\end{equation}
or, substituting the components of the $3\times 3$ matrix explicitly,
\begin{equation}\label{fresnel2}
  {\cal W} \stareq
  \frac{1}{3!}\hat{\epsilon}_{abc}\,\hat{\epsilon}_{def}\,
  \widetilde{\chi}^{\hat{0}a\hat{0}d}
  \,\widetilde{\chi}^{\hat{0}b\hat{0}e}
  \,\widetilde{\chi}^{\hat{0}c\hat{0}f}\stareq 0\,.
\end{equation}
Quite remarkably, we can rewrite the last equation as a 4D-covariant
equation.  Indeed, because of ${\hat{\epsilon}_{ a b c}\equiv
  \hat{\epsilon}_{\hat 0 a b c}}$, after some algebra, we find
\begin{equation}\label{wgen} 
  {\cal W}=\frac{\theta^2}{4!}\,\hat{\epsilon}_{mnpq}\,
  \hat{\epsilon}_{rstu}\, {\widetilde{\chi}}{}^{\,mnri}\,
  {\widetilde{\chi}}{}^{\,jpsk}\,{\widetilde{\chi}}{}^{\,lqtu } \,
  q_iq_jq_kq_l\,.
\end{equation} 
Here $\theta: = \det(e_i{}^\alpha)$. Since, by assumption, the 1-forms
$\vt^\a$ constitute a basis of the cotangent space, $\theta$ is always
nonvanishing.

Let us now define 4th-order {\bf T}amm--{\bf R}ubilar tensor density 
of weight $+1$,
\begin{equation}\label{G4}   
  {\cal G}^{ijkl}(\chi):=\frac{1}{4!}\,\hat{\epsilon}_{mnpq}\, 
  \hat{\epsilon}_{rstu}\, {\chi}^{mnr(i}\, {\chi}^{j|ps|k}\, 
  {\chi}^{l)qtu }\,.
\end{equation} 
In $n$ dimensions a totally symmetric tensor of rank $p$ has
$\binom{n+p-1}{p} =\binom{n-1+p}{n-1}$ independent components.
Accordingly, in four dimensions the TR-tensor (\ref{G4}) has 35
independent components. Because $\chi^{ijkl}=\widetilde\chi^{ijkl}+
\a\,\epsilon^{ijkl}$, the total antisymmetry of the Levi-Civita
$\epsilon$ evidently yields ${\cal G}(\chi)={\cal G}(\widetilde\chi)$.
Thus, starting from (\ref{wgen}) and discarding the irrelevant
nonvanishing factor $\theta^2$, we arrive at the generally covariant
4-dimensional (extended) Fresnel equation
\begin{equation} \label{Fresnel}     
{\cal G}^{ijkl}(\chi)\,q_i q_j q_k q_l = 0 \,.
\end{equation} 
This {quartic} equation in the components $q_i$ of the wave covector
$q$ was derived from a determinant of a $3\times 3$ matrix.
Apparently, the wave covectors $q$ lie on a {\em quartic Fresnel wave
  surface} in general, which, incidentally, is not exactly what we are
observing in vacuum at the present epoch of our universe.

\subsection{Decomposing Fresnel equation into time and space pieces}
\label{FresnelDec}

Physically, the zeroth component component of the wave covector is
interpreted as the frequency of a ``photon'', whereas the spatial 3-vector
part represents its momentum. Accordingly, the physical contents of
the Fresnel equation becomes more transparent after we carefully
separate it into its time and its space pieces. It is convenient to
denote the independent components of the TR-tensor (\ref{G4}) as
follows:
\begin{eqnarray}  
 \label{ma0} M &:=& {\cal G}^{0000} = \det{\cal A} \,,\\
M^a &:=& 4\,{\cal G}^{000a} = -\hat{\epsilon}_{bcd}\left( {\cal A}^{ba}
\,{\cal A}^{ce}\,{\cal C}^d_{\ e} + {\cal A}^{ab}\,{\cal A}^{ec}
\,{\cal D}_e^{\ d}\right)\,,\label{ma1}\\
 M^{ab} &:=& 6\,{\cal G}^{00ab} = \frac{1}{2}\,{\cal A}^{(ab)}\left[
({\cal C}^d{}_d)^2 + ({\cal D}_c{}^c)^2 - ({\cal C}^c{}_d + {\cal D}_d{}^c)
({\cal C}^d{}_c +  {\cal D}_c{}^d)\right]\nonumber\\ 
&&\qquad + ({\cal C}^d{}_c + {\cal D}_c{}^d)({\cal A}^{c(a}{\cal C}^{b)}{}_d 
+ {\cal D}_d{}^{(a}{\cal A}^{b)c}) - {\cal C}^d{}_d{\cal A}^{c(a}{\cal C}^{b)}
{}_c\nonumber\\ &&\qquad - {\cal D}_c{}^{(a}{\cal A}^{b)c}{\cal D}_d{}^d -
  {\cal A}^{dc}{\cal C}^{(a}{}_c {\cal D}_d{}^{b)} + 
  \left({\cal A}^{(ab)}{\cal A}^{dc}- {\cal A}^{d(a}{\cal A}^{b)c}\right)
   {\cal B}_{dc}\,,\label{ma2}\\
M^{abc} &:=& 4\,{\cal G}^{0abc} =
  \epsilon^{de(c|}\left[{\cal B}_{df}( {\cal A}^{ab)}\,{\cal D}_e^{\ f} 
 - {\cal D}_e^{\ a}{\cal A}^{b)f}\,) \right. \nonumber \\
&&\qquad\left.+ {\cal B}_{fd}({\cal A}^{ab)}\,{\cal C}_{\ e}^f - {\cal A}^{f|a}
{\cal C}_{\ e}^{b)}) +{\cal C}^{a}_{\ f}
\,{\cal D}_e^{\ b)}\,{\cal D}_d^{\ f} + {\cal D}_f^{\ a}
\,{\cal C}^{b)}_{\ e}\,{\cal C}^{f}_{\ d} \right] \,,\label{ma3}\\
M^{abcd} &:=& {\cal G}^{abcd} = \epsilon^{ef(c}\epsilon^{|gh|d}\,{\cal B}_{hf}
  \left[\frac{1}{2} \,{\cal A}^{ab)}\,{\cal B}_{ge} - {\cal C}^{a}_{\ 
      e}\,{\cal D}_g^{\ b)}\right] \,.\label{ma4}
\end{eqnarray}
Then the Fresnel equation (\ref{Fresnel}) in decomposed form reads
\begin{equation}
  q_0^4 M + q_0^3q_a\,M^a + q_0^2q_a q_b\,M^{ab} + q_0q_a q_b
  q_c\,M^{abc} + q_a q_b q_c q_d\,M^{abcd}=0\,.\label{decomp}
\end{equation} 

\section{Reducing the quartic wave surface to the light cone}\label{reduce}

\subsection{Maxwell-Lorentz electrodynamics in vacuum}\label{ML}

As a first example, let us demonstrate that our formalism yields the 
correct result for the conventional Maxwell-Lorentz spacetime relation 
with the constitutive tensor density
\begin{equation}\label{MLcon}
  \chi^{ijkl}=2\lambda_0\sqrt{-g}g^{i[k|}g^{j|l]}=
  \lambda_0\sqrt{-g}\left( g^{ik}g^{jl}-g^{il}g^{jk} \right)\,.
\end{equation}
Here $\lambda_0 = \sqrt{\varepsilon_0/\mu_0}$ is the so-called vacuum
impedance. A spacetime metric $g_{ij}$ is assumed on the manifold.
Substituting (\ref{MLcon}) into (\ref{G4}), we can calculate the
corresponding TR-tensor density straightforwardly:
\begin{equation} 
  {\cal G}^{ijkl} =
  -\,\lambda_0^3\sqrt{-g}\,g^{(ij}g^{kl)}=\frac{\lambda_0^3}{3}\sqrt{-g}
  \left(g^{ij}g^{kl}+ g^{kj}g^{il}+g^{lj}g^{ik} \right)\,.
\end{equation}
As a result, the quartic wave surface (\ref{Fresnel}) reduces to the
usual light cone (twice):
\begin{equation}
{\cal G}^{ijkl}q_iq_jq_kq_l = -\,\lambda_0^3\sqrt{-g}
\,(g^{ij}q_iq_j)(g^{kl}q_kq_l)=0.
\end{equation}

\begin{figure}
\epsfxsize=\hsize 
\epsfbox{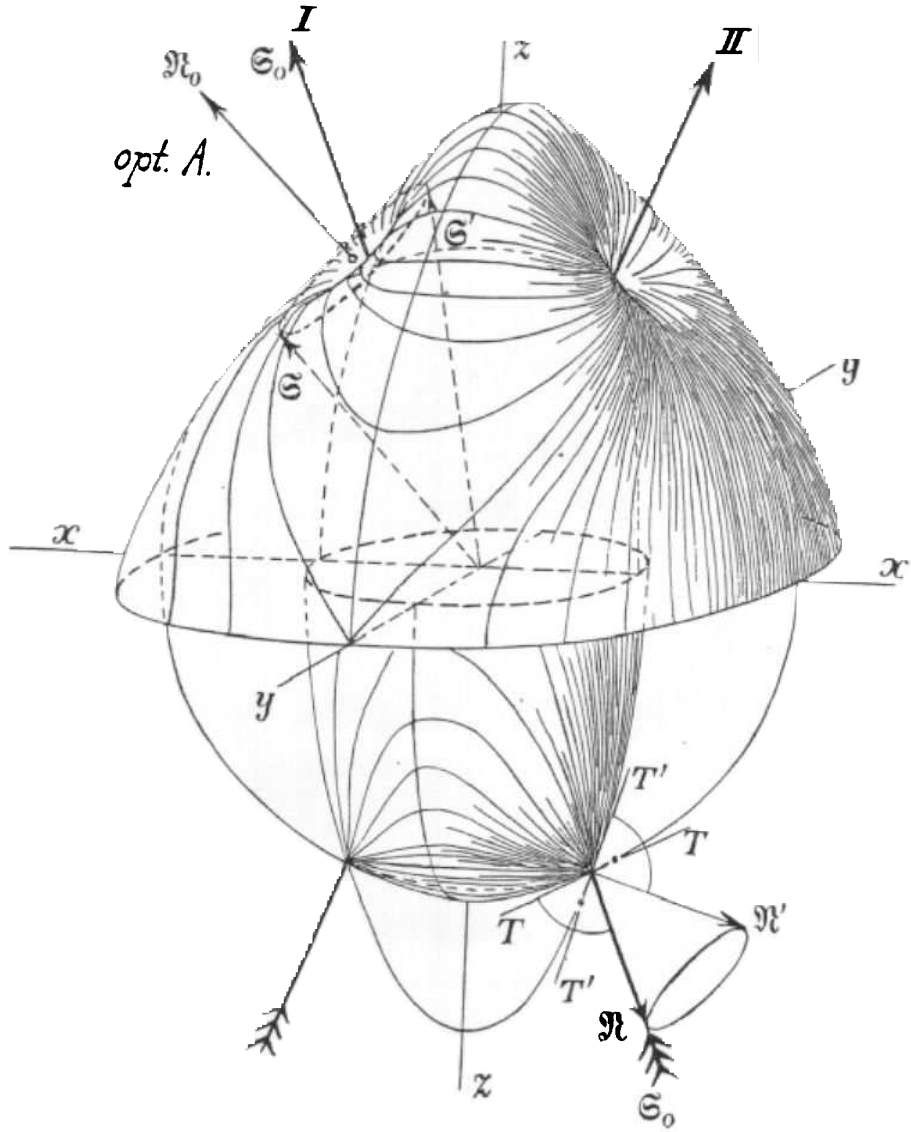}
\caption[{\it Clemens Schaefer's drawing\/} \cite{Schaefer32} of a quartic 
ray surface for $-{\cal A}\equiv\varepsilon=\varepsilon^T$, ${\cal
  B}\equiv \mu^{-1}=1$, and ${\cal C}=0$, ${\cal D}=0$: One ray vector
has 4 in general different cuts with the quartic surface. The Fresnel
equation (\ref{decomp1}), which describes the case under
consideration, defines a quartic {\it wave\/} surface that is dual to
the {\it ray\/} surface of this figure. Accordingly, both surfaces
encompass the same information and look, in fact, fairly similar.
Corresponding details are discussed in the Appendix.]
{\label{schaefer}{\it Clemens Schaefer's drawing\/} \cite{Schaefer32}
  of a quartic ray surface for $-{\cal
    A}\equiv\varepsilon=\varepsilon^T$, ${\cal B}\equiv \mu^{-1}=1$,
  and ${\cal C}=0$, ${\cal D}=0$: One ray vector has 4 in general
  different cuts with the quartic surface. The Fresnel equation
  (\ref{decomp1}), which describes the case under consideration,
  defines a quartic {\it wave\/} surface that is dual to the {\it
    ray\/} surface of this figure. Accordingly, both surfaces
  encompass the same information and look, in fact, fairly similar.
  Corresponding details are discussed in the Appendix. }
\end{figure}

\subsection{Constitutive tensor with vanishing electric/magnetic
  cross-terms}\label{cross}

The matrices ${\cal D}_b{}^a$ and ${\cal C}^b{}_a$ in the 
general spacetime relation (\ref{CR}) describe the mixing of 
electric and magnetic fields. Let us consider the case when these
cross-terms are absent, ${\cal C}^b{}_a = 0,\, {\cal D}_b{}^a = 0$, 
whereas ${\cal A}^{ab}$ and ${\cal B}_{ab}$ are left {\em a\/}symmetric. 
Physically, they are interpreted as the generalized 
permittivity and permeability tensors, respectively. Now the $M$-tensor 
densities (\ref{ma0})-(\ref{ma4}) simplify appreciably: $M^a =0$ and
$M^{abc} = 0$, while
\begin{eqnarray}\label{M}   
  M&=&\det{\cal A}\,,\\    M^{ab}&=&
  \left({\cal A}^{(ab)}{\cal A}^{dc}- {\cal A}^{d(a}{\cal
      A}^{b)c}\right){\cal B}_{dc}\,,\\ 
     M^{abcd}& =& \frac{1}{2}\,\epsilon^{ef(a}\epsilon^{|gh|b}
  \,{\cal A}^{cd)}\,{\cal B}_{ge}\,{\cal B}_{hf} \,.\label{M4}
\end{eqnarray}
The time and space decomposed Fresnel equation (\ref{decomp}) then reduces to
\begin{equation} \label{decomp1} 
  M\,q_0^4 + M^{ab}q_aq_b\,q_0^2 + M^{abcd}q_aq_bq_cq_d = 0.
\end{equation}
This bi-quadratic equation can be solved for the frequency square and 
yields a {\em Finsler\/} metric, see Rubilar \cite{Guillermo02},
\begin{equation}  
  \left(q_0^2+\frac{M^{ab}q_aq_b+ \sqrt{\Delta}}{2M}\right)
  \left(q_0^2+\frac{M^{ab}q_aq_b - \sqrt{\Delta}}{2M} \right) = 0\,,
\end{equation}
with $\Delta:={(M^{ab}q_aq_b)^2-4M\,M^{abcd}q_aq_bq_cq_d}$. A unique
light cone can be recovered, provided $\Delta = 0$ or
\begin{equation}
4 M M^{abcd} = M^{(ab}M^{cd)}.\label{cond1}
\end{equation} 
If this  sufficient condition for the existence of a light cone is fulfilled, 
then the quartic Fresnel wave surface reduces to
\begin{equation}
  \left(q_0^2+\frac{M^{ab}q_aq_b}{2M} \right)^2=0\qquad\text{ or}\qquad
  \left(g_{\rm opt}^{ik}\,q_iq_k\right)^2=0\,.
\end{equation}
Here the optical metric is introduced with the components $g_{\rm opt}^{00}
= 2M,\, g_{\rm opt}^{ab} = M^{ab}$. 

If the algebraic condition (\ref{cond1}) is not satisfied, the form of
the resulting quartic wave surface is extremely complicated. For
example, even in case when ${\cal B}_{ab} \sim \delta_{ab}$ and
${\cal A}^{ab} = {\cal A}^{ba}$ (which describes also the
light propagation in an anisotropic purely dielectric crystal), the
quartic Fresnel surface still looks highly nontrivial, see Fig.~2.

\subsection{Sufficient condition for unique light cone: 
electric/magnetic reciprocity}

The algebraic condition (\ref{cond1}) imposes a highly nontrivial
constraint on the components of the constitutive matrices. Although a
particular solution (for the homogeneous dielectric/magnetic medium)
can be easily derived, the general solution of (\ref{cond1}) is unknown.
There exists, however, another sufficient condition for the reduction
of the quartic Fresnel surface to a unique light cone which admits
a complete solution. Let us put skewon and axion fields to zero
$\!\not\!S_i{}^j=0,\,\a=0$. Then the spacetime relation (\ref{crypto2a}) 
contains only the first term on the right-hand-side with (\ref{CR})
constructed of the symmetric matrices ${\cal A}^{ab}={\cal A}^{ba}$,
${\cal B}_{ab} = {\cal B}_{ba}$ and the traceless matrices 
${\cal D}_a{}^b = {\cal C}^b{}_a$, with ${\cal C}^c{}_c =0$. 

Since the times of Maxwell and Heaviside, in the equations of 
electrodynamics a certain symmetry was noticed between the electric 
and the magnetic quantities and was used in theoretical discussions. 
We formulate electric/magnetic reciprocity as follows \cite{birkbook}: 
the energy-momentum current (\ref{simax}) is electric/magnetic reciprocal, 
i.e., it remains invariant $\Sigma_\alpha\rightarrow\Sigma_\alpha$ under 
the {transformation} 
\begin{equation}\label{duality1}  
  H\rightarrow \zeta F\,,\qquad F\rightarrow -\frac{1}{\zeta}\,H\,,
\end{equation}
with the twisted zero-form (pseudo-scalar function) $\zeta=\zeta(x)$
of dimension $[\zeta]=[H]/[F]=q^2/h$ = 1/resistance.  

We now require the spacetime relation also to be electric/magnetic
reciprocal. Then
\begin{equation}
  \zeta F_I=\,^{(1)}\kappa_I{}^K\left(-\frac{1}{\zeta}H_K\right)\qquad 
{\rm or}\qquad  -\zeta^2F_I=\,^{(1)}\kappa_I{}^K \,^{(1)}\kappa_K{}^LF_L.
\end{equation} 
Consequently, the constitutive tensor satisfies a {\it closure\/}
relation:
\begin{equation}\label{closure}
  ^{(1)}\kappa_I{}^K \,^{(1)}\kappa_K{}^L=-\zeta^2\delta_I^L.
\end{equation} 
Therefrom we find $\zeta^2=-\frac
16\text{Tr}(^{(1)}\kappa^2)=:\lambda^2$, and hence
$\left({}^{(1)}\kappa\right)^2 =-\lambda^2\mathbf{1}_6$. Thus,
symbolically, we can write
${}^{(1)}\kappa=\lambda\sqrt{-\mathbf{1}_6}$. With
$J:=\sqrt{-\mathbf{1}_6}$, we can then define the {\em duality operator}
\begin{equation}\label{dualityoperator}
  {}^\#:=J=\frac{1}{\lambda}\,^{(1)}\kappa.
\end{equation}
Besides the closure property $J^2=-\mathbf{1}_6$, this operator is
also symmetric by construction, $J(\phi)\wedge\psi=\phi\wedge
J(\psi)$, for all 2-forms $\phi$ and $\psi$. As a result, the
spacetime relation reads:
\begin{equation}
H=\lambda\,^\#F\quad\text{with}\quad [\lambda]=1/\text{resistance}\,.
\end{equation} 

Using the $3\times 3$-matrix parametrization (\ref{CR}) of the
constitutive tensor, we can solve the closure relation explicitly. In
matrix notation, $J=\begin{pmatrix}{\cal C} &{\cal B}&\\ {\cal
    A}&{\cal C}^T\end{pmatrix}$. Thus, the closure relation $J^2=-1$
reads:
\begin{equation}
{\cal C}^2 + {\cal A}{\cal B} = -1,\quad {\cal B}{\cal C} + {\cal C}^T
{\cal B} = 0,\quad {\cal C}{\cal A} + {\cal A}{\cal C}^T = 0.
\end{equation}
The solution is obtained straightforwardly in terms of ${\cal A}$ and an
arbitrary skew-symmetric matrix $\widehat{K} = - {\widehat K}{}^T$:
\begin{equation}
{\cal B} = - {\cal A}^{-1}\left[1 + (\widehat{K}{\cal A})^2\right],
\qquad {\cal C} = \widehat{K}{\cal A}^{-1}.
\end{equation}
Now we can substitute this solution into (\ref{ma0})-(\ref{ma4}) and find
the TR tensor density:
\begin{eqnarray}
M &=& {\rm det}{\cal A},\\
M^a &=& 4\widehat{k}{}^a,\\
M^{ab} &=& -2{\cal A}^{ab} + 6\widehat{k}{}^a
\widehat{k}{}^b/{\rm det}{\cal A},\\
M^{abc} &=& 4\left(-\,{\cal A}^{(ab}\widehat{k}{}^{c)} + 
\widehat{k}{}^a\widehat{k}{}^b\widehat{k}{}^c/{\rm det}{\cal A}
\right)/{\rm det}{\cal A},\\
M^{abcd} &=& \left({\cal A}^{(ab}{\cal A}^{cd)} - 2{\cal A}^{(ab}\widehat{k}
{}^c\widehat{k}{}^{d)}/{\rm det}{\cal A} + \widehat{k}{}^a\widehat{k}{}^b
\widehat{k}{}^c\widehat{k}{}^d/{\rm det}{\cal A}\right)/{\rm det}{\cal A}.
\end{eqnarray}
Here we denote $\widehat{k}{}^a := {\cal
  A}^{ab}\epsilon_{bcd}\widehat{K} {}^{cd}$. Finally, substituting
this into (\ref{decomp}), we can verify that the Fresnel equation reduces
to
\begin{equation}
  \left(g^{ik}q_iq_k \right)^2 = 0.
\end{equation}
Here the spacetime metric is constructed from the components of the 
constitutive matrices as follows
\begin{equation}
  g^{ik}=\frac{1}{\sqrt{-\det{\cal A}}}\begin{pmatrix}\det{\cal A}
 &\widehat{k}^b\\ \widehat{k}^a&-\,{\cal A}^{ab} + \widehat{k}^a
 \widehat{k}^b/\det{\cal A}\end{pmatrix}.
\end{equation}
It is not difficult to prove that this metric has Lorentzian signature
for every matrix ${\cal A}^{ab}$.

\section{Discussion and conclusion}

The main goal of this paper was to demonstrate that the light cone can
be recovered from a local and linear spacetime relation of classical
electrodynamics. Closure and symmetry of this spacetime relation are
sufficient conditions that guarantee the reduction of the general
quartic Fresnel wave surface to a unique light cone. The closure
relation alone, without the assumption of symmetry, is not sufficient
for the recovery of the light cone structure \cite{nonsym32}.

As a result of the reduction of the quartic Fresnel surface, we find
the spacetime metric with Minkowski (aka Lorentz) signature. This is
intimately related to the minus sign in the reciprocity transformation
(\ref{duality1}) and the closure relation (\ref{closure}). A plus
sign would yield the wrong Euclidean signature. Our approach shows that one
can treat the duality operator $\#$ as a metricfree predecessor of
the Hodge operator $\star$ that appears in the standard
Maxwell-Lorentz spacetime relation: $\# \;{\rm (duality\; operator)}
\; {\rm of \; Eq.\ (\ref{dualityoperator})}\longrightarrow \; \star\;
{\rm (Hodge \; operator)}\; {\rm of \; Eq.\ (\ref{constit1})}$.

Summarizing, the conformal part of the metric, that is, the light
cone, naturally emerges in our framework from a local and linear
spacetime relation.  In this sense, the light cone (and the spacetime
metric) is an {\em electromagnetic construct.}

\subsection*{Acknowledgments} 
Different versions of this paper have been given as seminars in
Vienna, Colum\-bia/Mis\-souri, Jena, and Bath. Respective discussions
with W.~Kummer, B.~Mashhoon, G.~Neugebauer \& A.~Wipf, and with
D.~Vassiliev are gratefully acknowledged. We also thank Yakov Itin and
Claus L\"ammerzahl for most helpful remarks. Our work has been
supported by the DFG project HE 528/20-1.

\section{Appendix: Wave and ray surfaces}

The drawing in Fig.\ 2 doesn't depict the {\it wave\/} surface but
rather the so-called {\it ray\/} surface which is dual to the wave
surface. In the particular case of Fig.\ 2, the propagation in an
anisotropic dielectric medium is discussed with ${\cal A}^{ab} =
-\,{\frac 1c}\,\varepsilon^{ab}$, $\varepsilon^{ab} = {\rm
  diag}(\varepsilon_1, \varepsilon_2, \varepsilon_3)$, and ${\cal
  B}_{ab} = c\,\delta_{ab}$ (with $c$ as speed of light). Then we can
immediately verify {}from (\ref{M})-(\ref{M4}) that $M = -
\varepsilon_1\varepsilon_2 \varepsilon_3/c^3$, $M^{abcd} =
c\,\delta^{(ab}\varepsilon^{cd)}$, and
\begin{equation}
M^{ab} = {\frac 1c}\begin{pmatrix} \varepsilon_1(\varepsilon_2 +
\varepsilon_3)& 0 & 0\\
0 &\varepsilon_2(\varepsilon_1 + \varepsilon_3) & 0 \\ 0 & 0 &
\varepsilon_3(\varepsilon_1 + \varepsilon_2)\end{pmatrix}.
\end{equation}
As a result, the Fresnel equation for the wave surface (\ref{decomp1})
can be recast into the simple form
\begin{equation}\label{wavesurf}
{\frac {\varepsilon_1\,q_1^2}{c^2\vec{q}^2 - q_0^2\,\varepsilon_1}} +
{\frac {\varepsilon_2\,q_2^2}{c^2\vec{q}^2 - q_0^2\,\varepsilon_2}} +
{\frac {\varepsilon_3\,q_3^2}{c^2\vec{q}^2 - q_0^2\,\varepsilon_3}} = 0.
\end{equation}
Here $\vec{q}^2 = q_1^2 + q_2^2 + q_3^2$. In crystal optics one
usually introduces a {\it ray 4-vector\/} $s$ which is dual to the
wave covector $q$, i.e., $s\rfloor q = 0$, see Kiehn et al.\ 
\cite{KiehnFresnel}.  The physical meaning of the ray vector is as
follows: Its spatial part coincides with the Poynting vector $\vec{s}
= \vec{E}\times\vec{H}$, whereas its time component describes the ray
velocity.  Then from (\ref{wavesurf}) one can straightforwardly derive
the dual equation of the Fresnel ray surface for the components of $s
= \{s^0, s^1, s^2, s^3\}$:
\begin{equation}\label{raysurf}
{\frac {(s^1)^2}{c^2\vec{s}^2 - (s^0)^2\,\varepsilon_1}} +
{\frac {(s^2)^2}{c^2\vec{s}^2 - (s^0)^2\,\varepsilon_2}} +
{\frac {(s^3)^2}{c^2\vec{s}^2 - (s^0)^2\,\varepsilon_3}} = 0.
\end{equation}
As we can see, the quartic surfaces (\ref{wavesurf}) and
(\ref{raysurf}) look pretty much similar, and thus depicting one of
them in fact gives a good idea about the dual surface. In Fig.~2, we
see the {\it ray surface} with $x = s^0s^1/\vec{s}^2,\, y =
s^0s^2/\vec{s}^2,\, z = s^0s^3/\vec{s}^2$. The vectors
$\mathfrak{S}_0$, $\mathfrak{S}$, and $\mathfrak{S}'$ represent the
particular rays related to the corresponding Poynting vectors, whereas
$\mathfrak{N}_0$, $\mathfrak{N}$, and $\mathfrak{N}'$ represent the
wave fronts (propagating along the corresponding wave covectors $q$)
dual to them. The Poynting vector $\vec{s} = \vec{E}\times\vec{H}$
describes the energy flux density, the wave vector
$\vec{q}\sim\vec{D}\times\vec{B}$ the momentum density.

\end{document}